\documentclass[12pt,letter]{article}

\usepackage{amsmath, amsthm, amssymb}
\usepackage{amssymb,epsfig}

\usepackage{hyperref}
\hypersetup{bookmarks=true}

\renewcommand{\baselinestretch}{1}

\makeatletter

\gdef\@punct{.\ \ }  % Punctuation after run-in section heading
\def\@sect#1#2#3#4#5#6[#7]#8{%
  \ifnum #2>\c@secnumdepth
     \def\@svsec{}
  \else
     \refstepcounter{#1}\edef\@svsec{%
     \ifnum #2>0{{\csname the#1\endcsname}}.\fi%
    \hskip .5em}
  \fi
  \@tempskipa #5\relax
  \ifdim \@tempskipa>\z@
     \begingroup #6\relax
       \@hangfrom{\hskip #3\relax\@svsec}{\interlinepenalty \@M #8\par}
     \endgroup
     \csname #1mark\endcsname{#7}
     \addcontentsline{toc}{#1}{\ifnum #2>\c@secnumdepth\else
          \protect\numberline{\csname the#1\endcsname}\fi#7}
  \else
     \def\@svsechd{#6\hskip #3\@svsec #8\@punct\csname
#1mark\endcsname{#7}
     \addcontentsline{toc}{#1}{\ifnum #2>\c@secnumdepth \else
          \protect\numberline{\csname the#1\endcsname}\fi#7}}
  \fi
  \@xsect{#5}}

\def\@ssect#1#2#3#4#5{\@tempskipa #3\relax
  \ifdim \@tempskipa>\z@
     \begingroup #4\@hangfrom{\hskip #1}{\interlinepenalty \@M
#5\par}\endgroup
  \else \def\@svsechd{#4\hskip #1\relax #5\@punct}\fi
  \@xsect{#3}}

\def\qed{\hskip 3pt \hbox{\vrule width4pt depth2pt height6pt}}

\newtheorem{Lemma}{Lemma}

\newtheorem{Theorem}[Lemma]{Theorem}

\newtheorem{Corollary}[Lemma]{Corollary}
\newtheorem{Definition}[Lemma]{Definition}

\usepackage{calc}
\usepackage{tikz}
\usetikzlibrary{decorations.markings}
\tikzstyle{vertex}=[circle, draw, inner sep=0pt, minimum size=6pt]
\tikzset{->-/.style={decoration={
markings,
mark=at position #1 with {\arrow{>}}},postaction={decorate}}}

\begin{document}

\title{On some distributed scheduling algorithms for wireless networks with hypergraph interference models}

\author{Ashwin~Ganesan%
  \thanks{A. Ganesan is an Associate Professor at the International School of Engineering (INSOFE), INSOFE Education Private Limited, Mumbai, Maharashtra, India.
%53 Deonar House, Deonar Village Road, Mumbai-88, India; 
\texttt{ashwin.ganesan@gmail.com}.}
% A.G.~is the corresponding author.}
}

\date{}

\maketitle

% \vspace{-7.0cm}
% \begin{flushright}
% %   \texttt{\Statusstring}\\[1cm]
% \end{flushright}
% \vspace{+5.0cm}

\begin{abstract}
\noindent  
It is shown that the performance of the maximal scheduling algorithm in wireless ad hoc networks under the hypergraph interference model can be further away from optimal than previously known. The exact worst-case performance of this distributed, greedy scheduling algorithm is analyzed.
\end{abstract}

\bigskip
\noindent\textbf{Index terms} --- hypergraph interference models, wireless networks, admission control, fractional chromatic number, upper bounds, maximal scheduling, distributed algorithms

%\newpage
%---------------------------------------------------------------
%\vskip 0.3in
% \newpage
\tableofcontents

%\vspace{+0.5cm}
%-------------------------------------

\section{Introduction}

A long-standing open problem is to develop simple, distributed scheduling algorithms for wireless networks that are provably efficient.  Distributed mechanisms for admission control and scheduling have lower communication overhead and lesser complexity than optimal, centralized algorithms.  A distributed, greedy scheduling algorithm for wireless networks which has been well-studied in the literature is maximal scheduling; it is suboptimal, and it is known that its worst-case performance is characterized by the interference degree of the conflict graph  \cite{Chaporkar:Kar:Luo:Sarkar:2008}, \cite{Ganesan:2010}, \cite{Ganesan:WN:2014}.    

Modeling interference using hypergraphs instead of graphs helps capture certain complexities and increase system capacity (cf. \cite{Sarkar:Sivarajan:1998}, \cite{McEliece:Sivarajan:1994},  \cite{Zhang:Song:Han:2016}). In the present work, we extend the previous results in the literature on performance analysis of the maximal scheduling algorithm to the more general case of hypergraphs.  We define the so-called interference degree of a hypergraph, which characterizes the worst-case performance of maximal scheduling.

%---------------------------------------------------------------
\section{System Model}

A hypergraph $H$ (on $L$) is a pair $(L,\mathcal{E})$, where $L$ is a set and $\mathcal{E}$ is a collection of subsets of $L$. The set $L$ is called the ground set of the hypergraph, and each element of $\mathcal{E}$ is called an edge (or a hyperedge).  A graph is a special case of a hypergraph where each hyperedge contains exactly two elements.  A subset of vertices is an independent set in the hypergraph if it does not contain any hyperedge. 

Let $G=(V,L)$ be a wireless network, where $V$ is a set of nodes and $L = \{\ell_1,\ldots,\ell_N\}$ $(\ell_i \in V \times V)$ is a set of wireless links.  Due to wireless interference, links in the same vicinity cannot be simultaneously active.  The interference is modeled using a (conflict) hypergraph $H=(L,\mathcal{E})$, where $L$ is the ground set of the hypergraph and $\mathcal{E}$ consists of the set of all subsets $E \subseteq L$ such that $E$ is a minimal subset of links that cannot be simultaneously active due to interference.  For example, consider a wireless network $G=(V,L)$ consisting of three nodes and three links $L=\{\ell_1, \ell_2, \ell_3\}$ that form a triangle. It is possible for the interference to be such that any two of these three links can be simultaneously active, but if all three links are simultaneously active the interference is intolerable. In that case, $F=\{\ell_1, \ell_2, \ell_3\}$ is a minimal forbidden set of links and therefore is a hyperedge. The subset $F$ is ``minimal'' in the sense that no proper subset of $F$ is forbidden.  The maximal independent sets of this hypergraph $H=(L, \{F\})$ are exactly $\{\ell_1, \ell_2\}$, $\{\ell_2, \ell_3\}$, and $\{\ell_1, \ell_3\}$.  To understand how this hypergraph model $H$ achieves better throughput than any conflict graph model on the same vertex set $L$, observe that there does not exist a conflict graph on vertex set $L$ which has the same set of independent sets as $H$.  For example, consider the conflict graph $G_c = (L, L')$ where $L' = \{\ell_1 \ell_2\}$ contains a single edge. Then, all independent sets of this conflict graph are also independent sets in the hypergraph and hence can be simultaneously active; however, there exist subsets of links such as $\{\ell_1, \ell_2\}$ which can be simultaneously active but which are not independent sets of the conflict graph.  Hence, there exist independent sets of the hypergraph which are not independent sets of the conflict graph.  The conflict graph cannot be taken to be the empty graph, because all three links form an independent set in the empty graph but cannot be simultaneously active.    Conflict graph models are conservative and underutilize the network resources because they have fewer sets of independent sets \cite{Sarkar:Sivarajan:1998}, \cite{Li:Negi:2012}.

The admission control problem is now formally stated.  Let $\tau = (\tau(\ell): \ell \in L)$ be a link demand vector, where the quality-of-service requirement for each link is specified by the fraction $\tau(\ell)$ of each unit of time that link $\ell$ demands to be active.  An independent set of the hypergraph $H=(L,\mathcal{E})$ is a subset $J \subseteq L$ that does not contain any hyperedge \cite{Berge:1989}. Let $\mathcal{I}(H)$ denote the set of all independent sets of $H$. A schedule is a map $t: \mathcal{I}(H) \rightarrow \mathbb{R}_{\ge 0}$ that assigns a time duration $t(J)$ to each independent set $J$ of the hypergraph.  The total duration of the schedule $t$ is $\sum_{J \in \mathcal{I}(H)} t(J)$.  The schedule $t$ satisfies demand $\tau$ if $\sum_{J: \ell_i \in J} t(J) \ge \tau(\ell_i)$, for all $\ell_i \in L$.  A link demand vector $\tau$ is said to be \emph{feasible} if there exists a schedule of duration at most $1$ satisfying demand $\tau$.  Given a conflict hypergraph $H$ and link demand vector $\tau$, the admission control problem is to determine whether $\tau$ is feasible.  The present focus is on distributed mechanisms for admission control, wherein feasibility is determined using only localized information.  Also, the admission control and scheduling problems have been well-studied for the conflict graph model (cf. \cite{Hajek:1984}, \cite{Hajek:Sasaki:1988},  \cite{Chaporkar:Kar:Luo:Sarkar:2008},  \cite{Ganesan:ToN:2020}), and the present work extends these results to hypergraph models.

An equivalent formulation of the admission control problem is as follows. Given the conflict hypergraph $H=(L,\mathcal{E})$, with $L = \{\ell_1,\ldots,\ell_N\}$, let $\mathcal{I}(H) = \{I_1,I_2,\ldots,I_K\}$ be the set of all independent sets of $H$.  Define the $N \times K$ $0,1$ link-independent set incidence matrix $M = [m_{ij}]$ by $m_{ij} =1$ if and only if $l_i \in I_j$. Let $\tau$ by a link demand vector.  The fractional chromatic number of the weighted hypergraph $(H,\tau)$, denoted by $\chi_f(H,\tau)$, is defined to be the optimal value of the linear program: minimize $1^T t$ subject to $Mt \ge \tau, t \ge 0$.  
A link demand vector $\tau$ is said to be \emph{feasible} if $\chi_f(H,\tau) \le 1$. Let $P_I = P_I(H)$ denote the convex hull of the characteristic vectors of the independent sets of $H$.  Then, $P_I$ is exactly the set of all link demand vectors that are feasible.

The notation used is standard. In the sequel, $[i]$ denotes the set $\{1,2,\ldots,i\}$.  Given a conflict hypergraph $H=(L,\mathcal{E})$, $N_i$ denotes the set of neighbors of link $\ell_i$ and is defined as the set of other links $\ell_j$ such that $\ell_i$ and $\ell_j$ belong to the same hyperedge: 
$$N_i = \{ \ell_j \in L: \{\ell_i, \ell_j\} \subseteq E, \mbox{ for some } E \in \mathcal{E} \}.$$  The set of edges of $H$ that contain $\ell_i$ is denoted $H(\ell_i)$. 

%---------------------------------------------------------------
\section{Distributed scheduling algorithms} \label{sec:suff:cond:Delta}

In this section, sufficient conditions for a link demand vector to be feasible are given. The sufficient conditions for admission control given in this section can be extended to provide a feasible schedule when the demand vector is feasible.  Thus, the results below give distributed algorithms for both admission control and scheduling.  The first sufficient condition is an extension of the greedy coloring algorithm for  graphs to the case of hypergraphs.  A special case of this sufficient condition, obtained when the hypergraph is a graph, is the row constraints of \cite{Hamdaoui:Ramanathan:05},  \cite{Gupta:Musacchio:Walrand:07}. 

There are two parallel sequences of papers in the literature – one is \cite{Hamdaoui:Ramanathan:05},  \cite{Gupta:Musacchio:Walrand:07}, \cite{Ganesan:2010}, \cite{Ganesan:WN:2014}  and another is \cite{Chaporkar:Kar:Luo:Sarkar:2008}, \cite{Li:Negi:2012}.   More specifically, distributed, greedy (maximal) scheduling algorithms have been studied in \cite{Hamdaoui:Ramanathan:05},  \cite{Gupta:Musacchio:Walrand:07}, \cite{Ganesan:2010}, \cite{Ganesan:WN:2014}, and the purpose of the present paper is to generalize these results to the case of hypergraphs.  The other parallel sequence of \cite{Chaporkar:Kar:Luo:Sarkar:2008}, \cite{Li:Negi:2012} formulates the problem using the framework of arrival processes and fluid limits.  It should be possible to abstract the essential details from their proofs, but by giving a more complete treatment as has been done in the present section, the following objectives are achieved: (a) The results and their proofs are more accessible to the reader – only techniques such as induction and combinatorics are required, (b)   The results that are essentially new can often be stated in the simpler model, and so it makes sense to give an accessible version of the statement of the theorem and its proof (with just these essential details).  The results and proofs recalled in the present section would be of interest to many researchers whose interest is in such techniques.  There are many directions in which these results can be extended.  Theorem~\ref{thm:LiNegi:suff:cond:W} and Corollary~\ref{cor:suff:cond:Li:Negi:Delta} are essentially from \cite{Li:Negi:2012}, but the proofs given here are based on our system model. 

\begin{Lemma}
 Let $H = (L, \mathcal{E})$ by a conflict hypergraph and let $\tau$ be a link demand vector.  A sufficient condition for $\tau$ to be feasible is that 

 $$ \tau(\ell_i) + \sum_{E \in H(\ell_i)} \min_{\ell_j \in E - \{\ell_i\}} \tau(\ell_j) \le 1, \forall \ell_i \in L.$$

\end{Lemma}

\noindent \emph{Proof:}
By the given inequality for $\ell_1$, it is possible to schedule $\ell_1$, i.e. it is possible to assign to link $\ell_1$ a subset $J_1 =[0,\tau(\ell_1)] \subseteq [0,1]$ of total length $|J_1| = \tau(\ell_1)$.   Suppose $\ell_1,\ldots, \ell_i$ have already been scheduled.  It will be shown that $\ell_{i+1}$ can also be assigned a subset of $[0,1]$ such that not all links in a hyperedge are simultaneously active (except possibly at endpoints of subintervals).  

Let $\mathcal{F}$ denote the set of hyperedges $E \in \mathcal{E}$ such that $E$ contains $\ell_{i+1}$ and all other links in $E$ have previously been scheduled, i.e. $E$ satisfies $\ell_{i+1} \in E$ and $E \subseteq \{\ell_1,\ldots,\ell_{i+1} \}$.  For $E \in \mathcal{F}$, define the common time slots of all links in $E$ that have already been scheduled:
$$\delta(E) = \bigcap_{j: \ell_j \in E, j \le i} J_j.$$
Because the set of links in a hyperedge cannot be simultaneously active, it is necessary that $J_{i+1}$ be disjoint from $\delta(E)$, for each $E \in \mathcal{F}$.  Also, $|\delta(E)| \le \min_{\ell_j \in E: j \le i} \tau(\ell_j)$.  It follows from the given inequality for $\ell_{i+1}$ that $\tau(\ell_{i+1}) + \sum_{E \in \mathcal{F}} |\delta(E)| \le 1$.  Hence, $\ell_{i+1}$ can also be assigned a subset $J_{i+1} \subseteq [0,1]$ such that $|J_{i+1}| =\tau(\ell_{i+1})$. 
\qed

% As a consequence of the above lemma, one obtains the following upper bound on $\chi_f(H,\tau)$:
% 
% $$ \chi_f(H, \tau) \le 
% \max_{\ell_i \in L} \left\{ 
% \tau(\ell_i) + \sum_{E \in H(\ell_i)} \min_{\ell_j \in E - \{\ell_i\}} \tau(\ell_j) \right\}
% .$$

Given a hypergraph $H=(L,\mathcal{E})$ with link set $L = \{\ell_1, \ldots, \ell_N\}$, let $\mathcal{W}$ denote the set of all $N \times N$ real, symmetric matrices $W$ such that (1) $W_{ij} \in [0,1], \forall i,j \in [N]$, (2) $W_{ii}=0$ for all $i \in [N]$, $W_{ij} = 0$ if $j \notin N_i$, and (3) $\sum_{j \in E} W_{ij} \ge 1$ for all $i \in E$ and $E \in \mathcal{E}$. 

\begin{Lemma} \label{lem:preliminary:inequality}
 Let $H = (L, \mathcal{E})$ be a conflict hypergraph, let $W \in \mathcal{W}$, and let $\tau$ be a link demand vector. Fix $i \ge 1$. Suppose $J_j \subseteq [0,1]$ satisfies $|J_j| = \tau(\ell_j)$, for all $j \le i$. Let $\mathcal{F}$ be the collection of hyperedges $E$ of $H$ which contain $\ell_{i+1}$ and such that every link in $E$ has index at most $i+1$. For $E \in \mathcal{F}$, define $\delta(E) = \cap_{j: \ell_j \in E, j \le i} J_j$. Then,
 $$ \left| \bigcup_{E \in \mathcal{F}} \delta(E) \right| \le \sum_{j \ne i+1} \{ W_{i+1,j} \tau(\ell_j) \}.$$
\end{Lemma}

\noindent \emph{Proof:}
The proof is by induction on $|\mathcal{F}|$.  If $|\mathcal{F}|=1$, then 
\begin{equation*}
\begin{split}
|\delta(E_1)| & = \left|\bigcap_{j: j \le i, \ell_j \in E_1} J_j \right| \\
& \le \min_{j: j \le i, \ell_j \in E_1} \tau(\ell_j) \\
& \le \sum_{j: j \le i, \ell_j \in E_1} \{ W_{i+1,j} \tau(\ell_j) \}
\end{split}
\end{equation*}
where we have used the facts $|J_j| = \tau(\ell_j)$ for $j \le i$, and $\sum_{j: j \le i, \ell_j \in E_1} W_{i+1,j} \ge 1$ for $W \in \mathcal{W}$.  Now fix $|\mathcal{F}| = k \ge 2$, where $\mathcal{F} = \{E_1,\ldots, E_k\}$,
and assume the assertion holds for all smaller values of $|\mathcal{F}|$.  It suffices to prove 
$$ | \delta(E_1) \cup \cdots \cup \delta(E_k)| \le \sum_{j \ne i+1} \{W_{i+1,j} \tau(\ell_j) \}.$$

Define the modified time intervals and demands
$$J'_j := J_j - \delta(E_k), \forall j \in [N]$$
$$\delta'(E_r) := \bigcap_{j: j \le i, \ell_j \in E_r} J'_j, \forall r \in [k-1]$$  
$$\tau'(\ell_j) := \tau(\ell_j) - |\delta(E_k) \cap J_j|, \forall j \in [N].$$
Then, the left-hand side can be rewritten as
$$| \delta(E_1) \cup \cdots \cup \delta(E_k)| = | \delta'(E_1) \cup \cdots \cup \delta'(E_{k-1})| + |\delta(E_k)|.$$ 
Also, $|J'_j| = \tau'(\ell_j), \forall j \in [i]$. The right-hand side can be rewritten as
$$\sum_{j \ne i+1}\{W_{i+1,j} \tau(\ell_j) \}
= \sum_{j \ne i+1} \{W_{i+1,j} \tau'(\ell_j)\} + \sum_{j \ne i+1} \{W_{i+1,j} |\delta(E_k) \cap J_j|\}.$$
By the inductive hypothesis,
$$| \delta'(E_1) \cup \cdots \cup \delta'(E_{k-1})|
\le \sum_{j \ne i+1} \{W_{i+1,j} \tau'(\ell_j) \}.$$
It remains to show
$$ |\delta(E_k) \le \sum_{j \ne i+1} \{ W_{i+1,j} |\delta(E_k) \cap J_j|\}.$$
Because $\delta(E_k) \subseteq J_j$ for $\ell_j \in E_k - \{\ell_{i+1}\}$, one obtains
\begin{equation*}
\begin{split}
\sum_{j \ne i+1} \{W_{i+1,j} |\delta(E_k) \cap J_j| \} & \ge \sum_{j: j \le i, \ell_j \in E_k} \{W_{i+1,j} |\delta(E_k) \cap J_j|\}
\\ & \ge \sum_{j: j \le i, \ell_j \in E_k} \{W_{i+1,j} |\delta(E_k)| \}
\\ & \ge |\delta(E_k)|. 
\end{split}
\end{equation*}
\qed

\begin{Theorem} \label{thm:LiNegi:suff:cond:W} \cite[Theorem 1]{Li:Negi:2012}
Let $H=(L, \mathcal{E})$ be a conflict hypergraph and let $\tau$ be a link demand vector.  A sufficient condition for $\tau$ to be feasible is that there exists some $W \in \mathcal{W}$ such that $(I+W) \tau \le 1$.
\end{Theorem}

\noindent \emph{Proof:}
Let $\tau$ be a link demand vector and suppose there exists $W \in \mathcal{W}$ satisfying
$$ \tau(\ell_i) + \sum_{j \ne i} \{ W_{ij} \tau(\ell_j)\} \le 1, \forall i \in [N].$$
It needs to be shown that $\tau$ is feasible, i.e. that each link $\ell_i$ can be assigned a subset $J_i \subseteq [0,1]$ of total length $\tau(\ell_i)$ such that not all links in any hyperedge are assigned the same subinterval (except possibly for endpoints of subintervals).  Initially, when none of the links have been scheduled, $J_i = \phi$, for all $i \in [N]$.  By the inequality for $\ell_1$ given in the assertion, $\tau(\ell_1) \le 1$, whence link $\ell_1$ can be assigned the time interval $J_1 = [0, \tau(\ell_1)]$.  Suppose links $\ell_1,\ldots\ell_i$ have already been assigned time intervals $J_1,\ldots,J_i$, respectively.  It suffices to show that $\ell_{i+1}$ can also be scheduled.

In order to ensure that the time interval assigned to link $\ell_{i+1}$ does not conflict with those assigned already to its neighbors, consider the set of hyperedges $\mathcal{F} :=\{E \in \mathcal{E}: \ell_{i+1} \in E, \mbox{ and } j \le i+1, \forall \ell_j \in E\}$. In other words, it suffices to consider those hyperedges which contain the current link $\ell_{i+1}$ of interest and whose remaining links have already been scheduled; time intervals assigned to links $\ell_j$ with $j > i+1$ can be ignored at this time because $J_j = \phi$ for $j > i+1$.  

Define the common time slots $\delta(E) := \cap_{j: j \le i, \ell_j \in E} J_j$, for $E \in \mathcal{F}$.  To ensure that not all links in a hyperedge are simultaneously active, it is necessary and sufficient that $J_{i+1}$ be disjoint from $\delta(E)$ (except at endpoints of subintervals) for all $E \in \mathcal{F}$.   Hence, to show that the demand $\tau(\ell_{i+1})$ can be satisfied, it suffices to show that $\tau(\ell_{i+1}) + |\cup_{E \in \mathcal{F}} \delta(E)| \le 1$. By the inequality for $\ell_{i+1}$ given in the assertion, it suffices to prove 
$$ \left| \bigcup_{E \in \mathcal{F}} \delta(E) \right| \le \sum_{j \ne i+1} \{ W_{i+1,j} \tau(\ell_j) \}. $$  
But this follows from Lemma~\ref{lem:preliminary:inequality}.
\qed

A special case of an element of $\mathcal{W}$ is the matrix $\{\Delta_{ij}\}$ defined as follows. For $i \in [N]$, define
\[
\Delta_{ij} = 
\begin{cases}
 \max \left\{ \frac{1}{|E|-1}: E \in \mathcal{E},\{\ell_i, \ell_j\} \subseteq E \right\}, &\mbox{ if } j \in N_i \\
 0, & \mbox{ if } j \notin N_i
\end{cases}
\]

\begin{Corollary} \label{cor:suff:cond:Li:Negi:Delta} \cite{Li:Negi:2012}
Let $H=(L,\mathcal{E})$ be a conflict hypergraph and let $\tau$ be a link demand vector.  Define $\Delta_{ij} = \Delta_{ij}(H)$ as above.  Then, a sufficient condition for $\tau$ to be feasible is
$$
\tau(\ell_i) + \sum_{j \ne i} \{ \Delta_{ij} \tau(\ell_j)\} \le 1, \forall i \in [N].
$$
\end{Corollary}

\noindent \emph{Proof:}
  It can be verified that the matrix $D = [\Delta_{ij}]$ belongs to $\mathcal{W}$.  By Theorem~\ref{thm:LiNegi:suff:cond:W}, $\tau$ is feasible.
\qed

%---------------------------------------------------------------
\section{Worst-case Performance} \label{sec:worst:case:perf}

Corollary~\ref{cor:suff:cond:Li:Negi:Delta} gave a sufficient condition for admission control. This is equivalent to giving an upper bound on the fractional chromatic number $\chi_f(H,\tau)$.   Define
$$ B(H,\tau) = \max_{i \in [N]} \left\{ \tau(\ell_i) + \sum_{j \ne i} \left\{\Delta_{ij} \tau(\ell_j) \right\} \right\}.$$
It follows from Corollary~\ref{cor:suff:cond:Li:Negi:Delta} that $\chi_f(H,\tau) \le B(H,\tau)$. 
The worst-case performance of a sufficient condition is defined to be the largest factor by which the upper bound is away from optimal, and is defined by the hypergraph invariant 
$$\beta(H) := \sup_{\tau \ne 0} \frac{B(H,\tau)}{\chi_f(H,\tau)}.$$ 
Thus, $\beta(H) = \sup_{\tau \in P_I} B(H, \tau)$. In this section,  an analysis of the worst-case performance is carried out for the above sufficient condition. 

\begin{Definition}
Given a hypergraph $H=(L,\mathcal{E})$, define the following quantities for $i \in [N]$:
$$ \Delta_i' = \max_{J \subseteq N_i: J \in \mathcal{I}(H)} \sum_{j \in J} \Delta_{ij} $$

$$ \Delta_i'' = \max_{J \subseteq N_i: J \cup \{i\} \in \mathcal{I}(H)} 1+\sum_{j \in J} \Delta_{ij} $$

$$\Delta' = \max_{i \in [N]} \Delta_i'$$ 

$$\Delta'' = \max_{i \in [N]} \Delta_i''$$

Define the interference degree of a hypergraph to be
$$ \sigma(H) = \max\{\Delta', \Delta''\}.$$
\end{Definition}

\bigskip \noindent 
\emph{Remarks:} In \cite[Theorem 2]{Li:Negi:2012}, it is claimed that the worst-case performance of Corollary~\ref{cor:suff:cond:Li:Negi:Delta} is essentially characterized by $ \Delta = \max_{i \in [N]} \max\{ 1, \Delta_i' \}$.
It is claimed in \cite[Theorem 2]{Li:Negi:2012} that if a link demand vector $\tau$ is feasible, then $\tau/\Delta$ will satisfy the sufficient condition of Corollary~\ref{cor:suff:cond:Li:Negi:Delta};  however, there seems to be an error in their proof, and a counterexample is given below to show that the worst-case performance can be a factor of more than $\Delta$ away from optimal.

In the special case where the conflict hypergraph is a conflict graph, when a link $i$ is scheduled, none of its neighbors can be scheduled during the same time slot. In this special case, $\Delta_{ij} = 1$ if $j \in N_i$, and so $\Delta_i'$ is the largest size of an independent set in $N_i$.  The graph invariant $\Delta'$ is the interference degree or induced star number of the conflict graph (cf. \cite{Chaporkar:Kar:Luo:Sarkar:2008},  \cite{Ganesan:2010}, \cite{Ganesan:WN:2014}).  Also, in the graph case, $\Delta_i'' = 1$, for all $i \in [N]$. Thus, $\sigma(H)$ reduces to the induced star number $\sigma(G_c)$ of the conflict graph $G_c$ in this special case. 

However, when generalizing this analysis to hypergraphs, a crucial difference needs to be taken into account.  In a hypergraph, when focusing on a particular link $\ell_i$ and maximizing the resource estimate $\tau(\ell_i) + \sum_{j \ne i} \Delta_{ij} \tau(\ell_j)$, two different cases can arise for a particular time slot.  In the first case, a link $\ell_i$ is not scheduled because all of its neighbors in some hyperedge containing it are already scheduled.  In this case, the resource estimate is at most $\Delta_i'$.  In the second case, a link $\ell_i$ can be scheduled at the same time as some of its neighbors - for example, in the example hypergraph below, link $1$ can be scheduled at the same time as links $2, 3, 5, 6$ because $\{1,2,3,5,6\}$ is an independent set of the hypergraph.  In such cases, the maximum contribution to the resource estimate during one unit of time can be as large as $\Delta_i''$.   Thus, $\Delta_i''$ can't be ignored for hypergraph models; it appears that the proof of \cite[Theorem 2]{Li:Negi:2012} has overlooked the term $\Delta_i''$ in the formula for the worst-case performance. 

\begin{Theorem}
There exist conflict hypergraphs $H=(L,\mathcal{E})$ and link demand vectors $\tau$ such that the sufficient condition of Corollary~\ref{cor:suff:cond:Li:Negi:Delta} can be away from optimal by a factor larger than $\Delta$.
\end{Theorem}

\noindent \emph{Proof:}
Let $H=(L,\mathcal{E})$ be the conflict hypergraph $\mathcal{E} = \{E_1, E_2\}$, where $E_1 = \{1,2,3,4\}$, $E_2 = \{1,5,6,7\}$.  Here, links are labeled as $i$ instead of $\ell_i$ for simplicity of exposition.  It needs to be proved that the worst-case performance $\beta(H) = \sup_{\tau \in P_I} B(H,\tau)$ is larger than $\Delta$. Consider the link demand vector $\tau = (\frac{1}{2}, 1, 1,$ $\frac{1}{2}, 1, 1, \frac{1}{2})$.  Then $\tau$ is feasible because there exists a schedule of duration at most $1$ satisfying $\tau$: the two independent sets $\{1,2,3,5,6\}$ and $\{2,3,4,5,6,7\}$ can each be active for duration $\frac{1}{2}$ units, and this gives a schedule satisfying $\tau$.  It will be proved that $B(H,\tau) > \Delta$.  It suffices to show that $B(H,\tau) \ge \frac{13}{6}$ and $\Delta=2$. Observe that $B(H,\tau) := \max_{i \in [N]} \left\{ \tau(\ell_i) + \sum_{j \ne i} \left\{\Delta_{ij} \tau(\ell_j) \right\} \right\}$ is at least $\tau(\ell_1)+\sum_{j \ne 1} \left\{\Delta_{1j} \tau(\ell_j) \right\}$ $= \frac{1}{2} + \frac{1}{3} (1+1+\frac{1}{2} + 1 + 1 + \frac{1}{2}) = \frac{13}{6}$.  Also, $\Delta_1' = \max \{ \sum_{j \in J} \Delta_{1j}: J \subseteq N_1, J \in \mathcal{I}(H) \}$ 
$ = \sum_{j \in \{2,3,4,5,6,7 \} } \Delta_{1j} = \frac{1}{3} \times 6 = 2$, and $\Delta_i'=1$ for $i=2,3,4$. Thus, $\Delta = \max_i \max\{1, \Delta_i'\} = 2$, as required. 
\qed

The demand vector given in the previous example is not uniform on the subset $\{2,3,\ldots,7\}$. However, it appears that the hypergraph $H$ has some kind of symmetry whereby the links in $\{2,3,4\}$ can be scheduled in a round-robin manner, and similarly for the links in $\{5,6,7\}$, so that the demand vector satisfied by such a schedule has a uniform demand pattern on the subset $\{2,3,\ldots,7\}$ and still attains the same value for the upper bound $B(H,\tau)$.  Indeed, the automorphism group of the hypergraph $H$ can be used to show that the worst-case performance is attained by a uniform demand pattern. 

An automorphism of a hypergraph $H=(L,\mathcal{E})$ is a permutation $\pi$ of $L$ that maps the set of hyperedges to itself.  The automorphism group of the hypergraph, denoted by $Aut(H)$, is the set of all automorphisms of $H$. In other words, $Aut(H) := \{ \pi \in Sym(L): \pi(\mathcal{E}) = \mathcal{E} \}$, where $Sym(L)$ is the full symmetric group acting on $L$.  Thus, if $\mathcal{E} = \{E_1, E_2\}$, where  $E_1 = \{1,2,3,4\}$, $E_2 = \{1,5,6,7\}$, then $Aut(H)$ fixes the point $1$ and is the automorphism group of the partition of a $6$-element set into two $3$-subsets, and hence is isomorphic to $(S_3 \times S_3) \rtimes \mathbb{Z}_2$ (cf. \cite[p. 46]{Dixon:Mortimer:1996}).  Also, $Aut(H)$ acts transitively on the set $\{2,3,\ldots,7\}$.

The exact value of the worst-case performance $\beta(H)$ is computed next (see also Corollary~\ref{cor:beta:star:perf} for another proof). 

\begin{Lemma} \label{lem:hyp:eg:7:by:3}
Let $H=(L,\mathcal{E})$ be the conflict hypergraph defined by $\mathcal{E} = \{ \{1,2,3,4\}$, $\{1,5,6,7\} \}$. Then, the worst-case performance  $\beta(H)$ of the sufficient condition of Corollary~\ref{cor:suff:cond:Li:Negi:Delta} is exactly $\frac{7}{3}$. 
\end{Lemma}

\noindent \emph{Proof:}
Let $B(H, \tau, i) = \tau(\ell_i) + \sum_{j \ne i} \{ \Delta_{ij} \tau(\ell_j) \}$. Then, $B(H, \tau) = \max_{i \in N} B(H, \tau, i)$, and the objective is to compute $\beta(H) = \sup_{\tau \in P_I} B(H,\tau)$.  If $i \ge 2$ and link $j$ belongs to a different hyperedge than link $i$, then $\Delta_{ij} = 0$, and so the value $B(H, \tau, i)$ can also be attained for some $i=1$ and a suitable $\tau$.  Thus, $\beta(H) = \sup_{\tau \in P_I} B(H, \tau, 1)$. 

Let $\tau$ be a demand vector and let $\pi \in Aut(H)$.  Let $\pi(\tau)$ denote the demand vector obtained by permuting the components of $\tau$ according to permutation $\pi$. Then, $\Delta_{1j} = \Delta_{1 \pi(j)}$ because an automorphism preserves the sizes of hyperedges.  It follows that $B(H, \tau, 1) = B(H, \pi(\tau), 1)$.  The convex combination $$\tau' = \frac{1}{|Aut(H)|} \sum_{\pi \in Aut(H)} \pi(\tau)$$ 
has a uniform demand pattern on $N_1 = \{2,3,\ldots,7\}$ because $Aut(H)$ acts transitively on this subset (cf. \cite[p. 5]{Scheinerman:Ullman:1997}). Also, $\tau'$ is feasible if $\tau$ is and $B(H, \tau', 1) = B(H, \tau, 1)$.  It follows that $\beta(H)$ is achieved for some $\tau$ that is uniform on $\{j: j \ne 1\}$.  

The hypergraph $H$ has $10$ maximal independent sets: $J_1, \ldots, J_9$ are subsets of the form $\{1\} \cup (E_1 - \{j_1\}) \cup (E_2 - \{j_2 \})$ and $J_{10} = (E_1 \cup E_2) - \{1\}$.  Consider a schedule that assigns a duration $a$ to each independent set $J_k$ ($k \in [9]$) and duration $b$ to $J_{10}$. Every demand vector uniform on $N_1$ is satisfied by a schedule of this form.  The demand pattern $\tau'$ satisfied by this schedule is $\tau'(1) = 9a$ and $\tau'(j) = 6a+b$ ($j \ge 2$).  Thus, $B(H, \tau', 1) = \tau'(1) + \sum_{j \ne i} \Delta_{ij} \tau(j) = 21a+2b$.  The duration of this schedule is $9a+2b$. Thus, $\beta(H)$ is the optimal value of the linear program: maximize $21a+2b$ subject to $9a+b \le 1$, $a,b \ge 0$. Evaluating the objection function at the three vertices of the feasibility polytope, one obtains $\beta(H) = 7/3$. 
\qed

%================================
\begin{Theorem} \label{thm:beta:equals:sigma}
Let $H=(L,\mathcal{E})$ be a conflict hypergraph.  The worst-case performance of the sufficient condition of Corollary~\ref{cor:suff:cond:Li:Negi:Delta} is given by the interference degree of the hypergraph, i.e. $$\beta(H) = \sigma(H)$$
\end{Theorem}

\noindent \emph{Proof:}
 First, it will be shown that $\beta(H) \le  \sigma(H) $.  Let $B(H, \tau, i) = \tau(\ell_i) + \sum_{j \ne i} \{ \Delta_{ij} \tau(\ell_j) \}$.  Let $\tau$ and $i$ be such that $\tau$ is a feasible demand vector and $\beta(H) = B(H, \tau, i)$. Let $t$ be a schedule satisfying $\tau$, and suppose $t$ assigns duration $t(I_k)$ to independent set $I_k \in \mathcal{I}(H)$.  The contribution to $B(H, \tau, i)$ due to demands satisfied during this $k$-th time slot is at most $t(I_k) \sigma(H)$ because the contribution is at most $t(I_k) \Delta_i'$ if link $i$ is not active during this time slot, and the contribution is at most $t(I_k) \Delta_i''$ if link $i$ is active during this time slot. Summing over the contributions to $B(H, \tau, i)$ during the entire duration $[0,1]$, one obtains that $B(H, \tau, i)  \le \sum_k t(I_k) \sigma(H) \le \sigma(H)$. This proves that $\beta(H) \le \sigma(H)$.  

To prove the opposite inequality, suppose that the values of $i$ and $J$ attaining $\sigma(H)$ are known; denote these values by $i$ and $J$, respectively.  Choose the demand pattern $\tau$ to be the characteristic vector of $J$ or $J \cup \{i\}$, according as whether $\Delta_i'$ or $\Delta_i''$ is larger. For this demand vector $\tau$, $B(H, \tau, i) = \sigma(H)$. Hence, $\beta(H) \ge \sigma(H)$.
\qed

Theorem 8 is tight because it characterizes the \emph{exact} worst case performance of \emph{every} network, i.e. it states that $\beta(H)$ equals $\sigma(H)$ for every $H$; so $\sigma(H)$ is not an upper or lower bound for $\beta(H)$ but the exact value. 
The proof of Theorem~\ref{thm:beta:equals:sigma} implies that a $0,1$-valued (characteristic vector) $\tau$ always achieves the supremum, and the proof of Lemma~\ref{lem:hyp:eg:7:by:3} shows there can also be other demand patterns that achieve the supremum. The demand pattern $\tau = (1,1,1,0,1,1,0)$ isn’t the only $0,1$-valued worst-case demand pattern – the $0$’s can occur in other positions by symmetry. A convex combination of these $0,1$-valued worst-case demand patterns is the worst-case demand pattern given in Lemma~\ref{lem:hyp:eg:7:by:3}.

A $\beta$-star of a hypergraph $H=(L,\mathcal{E})$ is a collection of edges $\mathcal{F} \subseteq \mathcal{E}$ satisfying the following condition: there exists some $x \in L$ such that $E \cap F = \{x\}$ for all $E, F \in \mathcal{F}, E \ne F$. A $\beta$-star is a generalization of the star $K_{1,r}$ subgraph found in graphs and satisfies the property that any two edges have exactly one vertex in common, and this vertex which is common to all the edges is the center of the star.  

\begin{Corollary} \label{cor:beta:star:perf}
Let $H$ be a $\beta$-star containing exactly $n_k$ edges of size $k$ $(k \ge 2)$.  Then, the worst-case performance $\beta(H)$ of the sufficient condition of Corollary~\ref{cor:suff:cond:Li:Negi:Delta} is given by
$$\beta(H) = \max\left\{ |\mathcal{E}|, 1 + \sum_k n_k \frac{k-2}{k-1} \right\}.$$
\end{Corollary}

\noindent \emph{Proof:}
By Theorem~\ref{thm:beta:equals:sigma}, $\beta(H) = \max\{\Delta', \Delta'' \}$. If $H$ is a $\beta$-star, then it can be verified that the value of $\Delta'$ is the number of edges in the hypergraph, and the value of $\Delta''$ is the second parameter in the assertion.
\qed

The proofs of Lemma~\ref{lem:hyp:eg:7:by:3} and Corollary~\ref{cor:beta:star:perf} both imply that for the specific hypergraph  $\mathcal{E} = \{ \{1,2,3,4\}$, $\{1,5,6,7\} \}$, the worst-case performance is $\beta(H)=\frac{7}{3}$.  However, the two proof techniques give different demand patterns $\tau$ that achieve this worst-case performance.  In the first proof, the worst-case demand pattern is the convex combination of $9$ independent sets, giving $\tau = (1, \frac{2}{3}, \ldots, \frac{2}{3})$, and so the support of $\tau$ is the set $L$ of all links. In the second proof, the worst-case demand pattern is of the form $\tau = (1,1,1,0, 1,1,0)$ and is the characteristic vector of an independent set.  

%---------------------------------------------------------------
\section{Concluding Remarks}
%====================================

It is important to characterize the worst-case performance of distributed admission control and scheduling algorithms because they can overestimate the network resources required to satisfy a given set of demands by up to this factor.   The interference degree of a hypergraph was defined and it was shown that in the worst case, the performance of the maximal scheduling algorithm is away from that of an optimal, centralized algorithm by a factor equal to the interference degree of the hypergraph. 

\section{Acknowledgements}

Thanks are due to the anonymous reviewers for helpful suggestions.

 {
\bibliographystyle{plain}
\bibliography{refs_ag.bib}

}
\end{document}